\documentclass[11pt]{article}
\usepackage{graphicx}
\usepackage{latexsym,graphicx}
\usepackage{float}
\usepackage{amsmath}
\usepackage{amsfonts}
\usepackage{amssymb}
\usepackage{epstopdf}
\usepackage{color}
\usepackage{cite}
\DeclareGraphicsExtensions{.eps}
\usepackage[left=2.5cm,top=2.5cm,right=2.5cm,bottom=2.5cm]{geometry}

\catcode `\@=11
\catcode `\@=12
\begin{document}
	\begin{center}
		\large{\bf{Constrained transit cosmological models in $f(R,L_{m},T)$-gravity}} \\
		\vspace{5mm}
		\normalsize{ Dinesh Chandra Maurya}\\
		\vspace{5mm}
		\normalsize{Centre for Cosmology, Astrophysics and Space Science, GLA University, Mathura-281 406,
		Uttar Pradesh, India.}\\
		\vspace{2mm}
		E-mail:dcmaurya563@gmail.com \\
	\end{center}
	\vspace{5mm}
	\begin{abstract}
     In the present paper, we investigate constrained transit cosmological models in the most recent proposed modified gravity theory, $f(R,L_{m},T)$-gravity. We obtain the modified field equations for a flat homogeneous and isotropic Friedmann-Lema\^{\i}tre-Robertson-Walker (FLRW) spacetime metric. We constrain the equation of continuity by imposing the equation of state for the perfect fluid source $p=-\frac{1}{3}\rho+p_{0}$ so that we get energy conservation equation as $\dot{\rho}+3H(\rho+p)=0$, (because generally, energy conservation law is not satisfied in $f(R,L_{m},T)$-gravity). Using this constraint, we establish a relation between the energy density parameters $\Omega_{m0}$, $\Omega_{r0}$, and $\Omega_{f0}$ and the Hubble function. After that, we made observational constraints on $H(z)$ to obtain the best-fit present values of $\Omega_{m0}$, $\Omega_{r0}$, and $H_{0}$. Then, we use these best-fit values of energy parameters to investigate cosmological parameters such as the deceleration parameter, the effective equation of state $\omega_{eff}$, and the energy density parameters $\Omega_{m}$, $\Omega_{r}$, and $\Omega_{f}$ to learn more about the components and history of the expanding universe. We found an effective EoS parameter in the range $-1\le \omega_{eff}\le\frac{1}{3}$ with a deceleration-acceleration transition redshift value of $z_{t}=0.6377, 0.6424$ along two datasets cosmic chronometer (CC) and Pantheon SNIa, respectively.
	\end{abstract}
	\smallskip
	\vspace{5mm}
	\vspace{1cm}
	
	\section{Introduction}
	It is well known from the observational results obtained in \cite{ref1,ref2,ref3,ref4,ref5} that the universe is currently passing through an accelerating phase of expansion. While General Relativity (GR) has proven to be the most successful theory for explaining the evolution of an expanding universe, it has been unable to adequately explain the current phase of the universe's accelerating expansion. In past decades of literature, cosmologists have attempted to explain this acceleration in the expansion of the universe in two different ways: first, by modifying the GR, and second, by presenting alternate gravity theory to the GR. Einstein explained this acceleration by inserting the cosmological constant $\Lambda$ in his field equation obtained from GR \cite{ref6}. The $\Lambda$CDM standard cosmological model of the universe can be understood as the best approximation between the predictions by the cosmological constant in GR and observations. The $\Lambda$CDM model also needed the presence of an extremely interesting but presently unknown component of the universe that is referred to as dark matter \cite{ref7}. This $\Lambda$CDM model is well fitted with the observed universe \cite{ref8,ref9,ref10,ref11}, yet it has several challenges that are to be theoretically justified.\\
	
	Several cosmologists have employed an alternative method to explain the acceleration of the universe's expansion, rather than altering the action. To understand the dynamical nature of dark energy, a review of various forms of dark energy models with observational constraints is presented in \cite{ref12}. The late-time accelerating phase of the universe is discussed by reconstructing the deceleration parameter motivated by thermodynamics in \cite{ref13}. Additional barotropic fluids describe the nature of dark energy and dark matter \cite{ref14}.  In \cite{ref15}, a model-independent approach to investigating dark energy and modified gravity problems via the cosmography technique is discussed on the various observational datasets. Repulsive gravity with curvature eigenvalues is discussed in \cite{ref16,ref17}. In \cite{ref18}, the principle of holographic dark energy and dark matter with second-order invariants has been used to investigate the speeding up of the expanding universe. In another approach, baryons and cold dark matter create pressure, which reflects the effects of cosmological constants \cite{ref19}. In \cite{ref20}, intermediate redshift calibration of gamma-ray bursts and cosmic constraints in non-flat cosmology are used to find cosmological correlations between different parameters.\\
	
	One can find several modified gravity theories in the literature, derived from altering the Lagrangian in the action, to explain the acceleration and various phenomena of the expanding universe. Some of these gravity theories are $f(R)$ gravity which was initially proposed in \cite{ref21} and developed in its early stages in \cite{ref22,ref23,ref24,ref25}. In \cite{ref26,ref27,ref28,ref29,ref30,ref31,ref32,ref33,ref34,ref35,ref36,ref37,ref38,ref39,ref40}, the cosmological and astrophysical properties of the universe models in $f(R)$ gravity have been comprehensively examined. The f(R) theory of gravity investigates the first cosmological model that describes both inflation and late-time acceleration in the evolution of the universe. An unification of the metric and the Palatini geometries extends the gravity theory as the hybrid metric-Palatini theory of gravity \cite{ref41,ref42,ref43,ref44}. Theories involving geometric structures that go beyond the Riemannian one, such as Weyl geometry, have also been investigated in \cite{ref45,ref46,ref47,ref48,ref49,ref50,ref51,ref52,ref53,ref54,ref55}. For reviews of modified gravity theories and their applications, see \cite{ref56,ref57,ref58,ref59,ref60,ref61,ref62,ref63,ref64,ref65,ref66,ref67,ref68,ref69}.\\
	
	To make cosmological models more general, \cite{ref70} created a new version of the gravitational theory in $f(R)$ by connecting matter and curvature from the outside in the form of $S=\int[f_{1}(R)+(1+\lambda f_{2}(R))L_{m}]\sqrt{-g}d^{4}x$. This extra coupling between matter and curvature creates extra forces. Harko \cite{ref71} then proposed a matter-geometry coupling modified gravity theory with the action $S=\int[\frac{1}{2}f_{1}(R)+G(L_{m}) f_{2}(R)]\sqrt{-g}d^{4}x$. This theory was later improved in \cite{ref72} as the $f(R, L_{m})$ gravity theory, where the Lagrangian function $f$ is any function of the Ricci curvature scalar $R$ and the matter Lagrangian $L_{m}$. After that, the basic ideas behind this matter-geometry coupling theory $f(R,L_{m})$ was studied in a number of ways \cite{ref73,ref74,ref75,ref76,ref77,ref78,ref79,ref80,ref81,ref82,ref83}. We have recently investigated some transit cosmological models in this $f(R,L_{m})$ gravity theory using constraints from observations in \cite{ref84,ref85,ref86,ref87,ref88}. After that, an alternate approach to matter-geometry coupling in the form of $f(R,T)$ gravity theory is presented in \cite{ref89}, where $R$ is the Ricci curvature scalar and $T$ is the trace of the energy-momentum tensor. Additionally, this geometry connects the trace of the matter-energy-momentum tensor in a non-minimal way. The impacts of $f(R,T)$ theories on cosmology and astrophysics are extensively discussed in \cite{ref90,ref91,ref92,ref93,ref94,ref95,ref96,ref97,ref98,ref99,ref100,ref101,ref102,ref103,ref104,ref105,ref106,ref107,ref108,ref109,ref110,ref111,ref112,ref113,ref114,ref115,ref116,ref117,ref118,ref119,ref120,ref121,ref122,ref123,ref124,ref125,ref126}.\\
	
	Recently, Haghani and Harko \cite{ref127} presented a generalized matter-geometry coupling theory called $f(R,L_{m},T)$ gravity. The unification of the modified gravity theories in Riemannian geometry, like the $f(R)$, $f(R,L_{m})$, and $f(R,T)$ gravity theories, is the main motivation behind the origin of this theory. Although theories $f(R)$-gravity, $f(T)$-gravity, and $f(Q)$-gravity may be somewhere interesting and effective to explain the evolution of the expanding universe, in this paper, we consider this generalized and less focused theory $f(R,L_{m},T)$ gravity to investigate the various cosmological properties of the expanding universe. This theory ensures that the modification is non-trivial for all kinds of matter fields. These are the physical motivations for considering this background in the study of cosmological models. A comparative study of a flat and a non-flat FLRW universe has been presented in \cite{ref128}. They showed that generalizing the models by adding a non-flat FLRW universe has important effects on the universe's overall dynamics, but it needs initial adjustments like whether it is open or closed, which are not needed for a flat FLRW universe. Here, we choose a flat FLRW spacetime to investigate the present model.\\
	
	The current paper's outline is as follows: In Section 1, there is a short introduction to the development of cosmological models in modified gravity theories on the Riemann manifold. There is also a short review of $f(R,L_{m},T)$ gravity theory and modified field equations in Section 2. In Section 3, we obtain some cosmological solutions for the particular function $f(R,L_{m},T)=R-\mu\,T\,L_{m}-\nu$. Here, $R$ is the Ricci curvature scalar, $L_{m}$ is the matter Lagrangian, $T$ is the trace of the energy momentum tensor $T_{ij}$, and $\mu$ and $\nu$ are the model-free parameters. In Sect. 4, we used MCMC analysis of cosmic chronometer (Hubble data) and Pantheon SNe Ia datasets to put some observational constraints on the cosmological parameters $H_{0}$, $\Omega_{m0}$, and $\Omega_{r0}$. In Sect. 5, we discuss the model results with recent findings, and in Sect. 6, we provide conclusions.
	
\section{$f(R,L_{m},T)$ gravity and field equations}

We consider the action for modified $f(R,L_{m},T)$ theory in the following form \cite{ref127}
\begin{equation}\label{eq1}
I=\frac{1}{16\pi}\int{f(R,L_{m},T)\sqrt{-g}d^{4}x}+\int{L_{m}\sqrt{-g}d^{4}x}
\end{equation}
where $f(R,L_{m},T)$ is an arbitrary function of the Ricci scalar, $R$, the trace $T$ of the stress-energy momentum tensor of the matter, $T_{ij}$, and of the matter Lagrangian density $L_{m}$. The function $f(R,L_{m},T)$ unifies two classes of theories of gravitation with matter-geometry coupling, namely, the $f(R,L_{m})$ gravity and $f(R,T)$ gravity theories. This theory ensures that the modification is non-trivial for all kinds of matter fields. These are the physical motivations for considering this background in the study of cosmological models. We define the stress-energy momentum tensor of the matter as \cite{ref129}
\begin{equation}\label{eq2}
  T_{ij}=-\frac{2}{\sqrt{-g}}\frac{\delta(\sqrt{-g}L_{m})}{\delta{g^{ij}}}
\end{equation}
and its trace by $T=g^{ij}T_{ij}$, respectively. By assuming that the Lagrangian density $L_{m}$ of matter depends only on
the metric tensor components $g_{ij}$, and not on its derivatives, we obtain
\begin{equation}\label{eq3}
  T_{ij}=g_{ij}L_{m}-2\frac{\partial{L_{m}}}{\partial{g^{ij}}}
\end{equation}
By varying the action $I$ of the gravitational field with respect to the metric tensor components $g^{ij}$ provides the following relationship
\begin{equation}\label{eq4}
  \delta{I}=\int\left[{f_{R}\delta{R}+f_{T}\frac{\delta{T}}{\delta{g^{ij}}}\delta{g^{ij}}+f_{L_{m}}\frac{\delta{L_{m}}}{\delta{g^{ij}}}\delta{g^{ij}}-\frac{1}{2}g_{ij}f\delta{g^{ij}}}-8\pi T_{ij}\delta{g^{ij}}\right]\sqrt{-g}d^{4}x,
\end{equation}
where we have denoted $f_{R}=\partial{f}/\partial{R}$, $f_{T}=\partial{f}/\partial{T}$ and $f_{L_{m}}=\partial{f}/\partial{L_{m}}$, respectively. For the variation of the Ricci scalar, we obtain
\begin{equation}\label{eq5}
  \delta{R}=\delta(g^{ij}R_{ij})=R_{ij}\delta{g^{ij}}+g^{ij}(\nabla_{\lambda}\delta\Gamma^{\lambda}_{ij}-\nabla_{j}\delta\Gamma^{\lambda}_{i\lambda})
\end{equation}
where $\nabla_{\lambda}$ is the covariant derivative with respect to the symmetric connection $\Gamma$ associated to the metric $g$. The variation of the Christoffel symbols yields
\begin{equation}\label{eq6}
  \delta{\Gamma^{\lambda}_{ij}}=\frac{1}{2}g^{\lambda\alpha}(\nabla_{i}\delta{g_{j\alpha}}+\nabla_{j}\delta{g_{\alpha i}}-\nabla_{\alpha}\delta{g_{ij}})
\end{equation}
and the variation of the Ricci scalar provides the expression
\begin{equation}\label{eq7}
  \delta{R}=R_{ij}\delta{g^{ij}}+g_{ij}\Box\delta{g^{ij}}-\nabla_{i}\nabla_{j}\delta{g^{ij}},
\end{equation}
Therefore, for the variation of the action of the gravitational field, we obtain
\begin{equation}\label{eq8}
 \delta{R}=\int\left[f_{R}R_{ij}+(g_{ij}\Box-\nabla_{i}\nabla_{j})f_{R}+f_{T}\frac{\delta(g^{\alpha\beta}T_{\alpha\beta})}{\delta{g^{ij}}}+f_{L_{m}}\frac{\delta{L_{m}}}{\delta{g^{ij}}}-\frac{1}{2}g_{ij}f-8\pi T_{ij}\right]\delta{g^{ij}}\sqrt{-g}d^{4}x
\end{equation}
We define the variation of $T$ with respect to the metric tensor as
\begin{equation}\label{eq9}
  \frac{\delta(g^{\alpha\beta}T_{\alpha\beta})}{\delta{g^{ij}}}=T_{ij}+\Theta_{ij},
\end{equation}
where
\begin{equation}\label{eq10}
  \Theta_{ij}=g^{\alpha\beta}\frac{\delta T_{\alpha\beta}}{\delta{g^{ij}}}=L_{m}g_{ij}-2T_{ij}
\end{equation}
for the perfect-fluid matter source.\\
Taking $\delta{I}=0$, we obtain the field equations of $f(R,L_{m}, T)$ gravity model as
\begin{equation}\label{eq11}
  f_{R}R_{ij}-\frac{1}{2}[f-(f_{L_{m}}+2f_{T})L_{m}]g_{ij}+(g_{ij}\Box-\nabla_{i}\nabla_{j})f_{R}=\left[ 8\pi+\frac{1}{2}(f_{L_{m}}+2f_{T})\right]T_{ij}.
\end{equation}
By contracting Eq.~(\ref{eq11}) gives the following relation between the Ricci scalar $R$ and the trace $T$ of the stress-energy tensor,
\begin{equation}\label{eq12}
f_{R}R-2[f-(f_{L_{m}}+2f_{T})L_{m}]+3\Box f_{R}=\left[ 8\pi+\frac{1}{2}(f_{L_{m}}+2f_{T})\right]T,
\end{equation}
where we have denoted $\Theta=\Theta_{i}^{i}$.\\
The problem of the perfect fluids,
described by an energy density $\rho$, pressure $p$ and four-velocity $u^{i}$ is more complicated, since there is no unique definition of the matter Lagrangian. However, in the present study we assume that the stress-energy tensor of the matter is given by
\begin{equation}\label{eq13}
  T_{ij}=(\rho+p)u_{i}u_{j}+pg_{ij},
\end{equation}
for the flat FLRW homogeneous and isopropic spacetime metric 
\begin{equation}\label{eq14}
	ds^{2}=-dt^{2}+a(t)^{2}(dx^{2}+dy^{2}+dz^{2})
\end{equation}
and the matter Lagrangian can be taken as $L_{m}=-\rho$. The four-velocity $u_{i}$ satisfies the conditions $u_{i}u^{i}=-1$ and $u^{i}\nabla_{j}u_{i}=0$, respectively. A comparative study of a flat and a non-flat FLRW universe has been presented in \cite{ref128}. They showed that generalizing the models by adding a non-flat FLRW universe has important effects on the universe's overall dynamics, but it needs initial adjustments like whether it is open or closed, which are not needed for a flat FLRW universe.\\
The energy conservation law can be obtained as
\begin{equation}\label{eq15}
	\dot{\rho}+3H(\rho+p)=\frac{(\rho+L_{m})(\dot{f}_{T}+\frac{1}{2}\dot{f}_{L_{m}})-\frac{1}{2}f_{T}(\dot{T}-2\dot{L}_{m})}
	{8\pi+f_{T}+\frac{1}{2}f_{L_{m}}}
\end{equation}

\section{Cosmological solutions for $f(R,L_{m},T)=R-\mu\,T\,L_{m}-\nu$}

To investigate the cosmological properties of the above-proposed modified gravity, we consider the following form of the arbitrary function $f(R,L_{m},T)$ (as suggested in \cite{ref127}).
\begin{equation}\label{eq16}
  f(R,L_{m},T)= R-\mu\,T\,L_{m}-\nu
\end{equation}
where $\mu$ and $\nu$ are the arbitrary positive constants.\\
Then
\begin{equation}\label{eq17}
	f_{R}=1,~~~~ f_{T}=-\mu\,L_{m}, ~~~~ f_{L_{m}}=-\mu\,T,
\end{equation}
By applying \eqref{eq16} in \eqref{eq11}, we get the following field equations:

\begin{equation}\label{eq18}
	R_{ij}-\frac{1}{2}R\,g_{ij}=(8\pi-\frac{\mu}{2}\,T-\mu\,L_{m})T_{ij}+(\mu\,L_{m}^{2}-\frac{\nu}{2})g_{ij}
\end{equation}
Now, using \eqref{eq13} and \eqref{eq14} in \eqref{eq18}, we obtain the following equations:

\begin{equation}\label{eq19}
	3H^{2}=8\pi\rho+\frac{\mu}{2}\rho^{2}-\frac{3\mu}{2}\rho p+\frac{\nu}{2}
\end{equation}\label{eq20}
\begin{equation}
	2\dot{H}+3H^{2}=-8\pi p-\frac{3\mu}{2}(\rho-p)p-\mu\rho^{2}+\frac{\nu}{2}
\end{equation}
From Eqs.~\eqref{eq15} and \eqref{eq16}, we obtain the equation of continuity as follows
\begin{equation}\label{eq21}
		\dot{\rho}+3H(\rho+p)=\frac{-\mu\rho(\dot{\rho}+3\dot{p})}{16\pi+3\mu(\rho-p)}
\end{equation}
Without loss of generality, we choose the equation of state (EoS) such that $\dot{\rho}+3H(\rho+p)=0$ and hence, from the right-hand side, we get $\dot{\rho}+3\dot{p}=0$ after integration, which gives the following equation of state:
\begin{equation}\label{eq22}
	p=-\frac{1}{3}\rho+p_{0}
\end{equation}
where $p_{0}$ is an integrating constant. Using this EoS, solving the equation of continuity, we obtain,
\begin{equation}\label{eq23}
	\rho=\rho_{m0}\left(\frac{a_{0}}{a}\right)^{2}-\frac{3}{2}p_{0}
\end{equation}
where $\rho_{m0}$ is the present value of matter energy density and $a_{0}$ is the present value of the scale factor $a(t)$.\\
Now, using Eqs.~\eqref{eq22} and \eqref{eq23} in \eqref{eq19}, we obtain the following expression
\begin{equation}\label{eq24}
	\Omega_{m}+\Omega_{r}+\Omega_{f}=1
\end{equation}
where $\Omega_{m}=\frac{(16\pi-9\mu p_{0})\rho_{m0}}{6H^{2}}\left(\frac{a_{0}}{a}\right)^{2}$ and $\Omega_{r}=\frac{\mu\rho^{2}_{m0}}{3H^{2}}\left(\frac{a_{0}}{a}\right)^{4}$ are matter and radiation energy density parameters, respectively, and $\Omega_{f}=\frac{\nu-24\pi p_{0}}{6H^{2}}$ is the dark energy density parameter due to $f(R,L_{m},T)$ function. From Eq.~\eqref{eq24}, one can obtain the following form of Hubble function:
\begin{equation}\label{eq25}
	H(t)=H_{0}\sqrt{\Omega_{m0}\left(\frac{a_{0}}{a}\right)^{2}+\Omega_{r0}\left(\frac{a_{0}}{a}\right)^{4}+\Omega_{f0}}
\end{equation}
Using the relation $\frac{a_{0}}{a}=1+z$, it can be expressed as
\begin{equation}\label{eq26}
	H(z)=H_{0}\sqrt{\Omega_{m0}(1+z)^{2}+\Omega_{r0}(1+z)^{4}+\Omega_{f0}}
\end{equation}
Now, we can define the effective energy density $\rho_{eff}$ and effective pressure $p_{eff}$ of the model as follows:
\begin{equation}\label{eq27}
	\rho_{eff}=\frac{16\pi\rho+\mu\rho^{2}-3\mu\rho p+\nu}{8\pi},~~~~p_{eff}=\frac{16\pi p+3\mu(\rho-p)p+2\mu\rho^{2}-\nu}{8\pi}
\end{equation}
Hence, the effective equation of state parameter $\omega_{eff}$ is obtained as
\begin{equation}\label{eq28}
	\omega_{eff}=-1+\frac{2\Omega_{m0}(1+z)^{2}+4\Omega_{r0}(1+z)^{4}}{3\Omega_{m0}(1+z)^{2}+3\Omega_{r0}(1+z)^{4}+3\Omega_{f0}}
\end{equation}
Now, using the Hubble function \eqref{eq26}, we derive the deceleration parameter as
\begin{equation}\label{eq29}
	q(z)=-1+\frac{\Omega_{m0}(1+z)^{2}+2\Omega_{r0}(1+z)^{4}}{\Omega_{m0}(1+z)^{2}+\Omega_{r0}(1+z)^{4}+\Omega_{f0}}
\end{equation}
The matter energy density parameter $\Omega_{m}$, radiation energy density parameter $\Omega_{r}$ and dark energy density parameter $\Omega_{f}$, are obtained as follows, respectively:
\begin{equation}\label{eq30}
	\Omega_{m}=\frac{\Omega_{m0}(1+z)^{2}}{\Omega_{m0}(1+z)^{2}+\Omega_{r0}(1+z)^{4}+\Omega_{f0}}
\end{equation}
\begin{equation}\label{eq31}
	\Omega_{r}=\frac{\Omega_{r0}(1+z)^{4}}{\Omega_{m0}(1+z)^{2}+\Omega_{r0}(1+z)^{4}+\Omega_{f0}}
\end{equation}
\begin{equation}\label{eq32}
	\Omega_{f}=\frac{\Omega_{f0}}{\Omega_{m0}(1+z)^{2}+\Omega_{r0}(1+z)^{4}+\Omega_{f0}}
\end{equation}

\section{Observational constraints}

This section contains a statistical analysis of our derived universe model parameters with observational datasets by applying MCMC (Monte Carlo Markov Chain) examination. To perform MCMC analysis, we use the emcee program, which is freely available at \cite{ref130}. We estimate the consistency of the solution in the model with two observational datasets: one is the cosmic chronometer (CC) data (Hubble data points), and the second is the pantheon SNIa datasets. These datasets pertain to the observed cosmos during a recent period.

\subsection{Hubble datasets $H(z)$}

We derive the Hubble function $H(z)$ from the model field equations in terms of redshift $z$ and model parameters that we are to determine from the analysis. In the present analysis, we consider 31 cosmic chronometer data points (Hubble data) \cite{ref131,ref132} in the range of redshift $0.07\le z \le 1.965$. These values were found using the differential ages (DA) of galaxies method. We can get a rough idea of the model parameters $H_{0}$, $\Omega_{m0}$, and $\Omega_{r0}$ by minimizing the $\chi^{2}$ function, which is the same thing as maximizing the likelihood function and is written as
\begin{equation}\nonumber
	\chi_{CC}^{2}(\phi)=\sum_{i=1}^{i=N}\frac{[H_{ob}(z_{i})-H_{th}(\phi, z_{i})]^{2}}{\sigma_{H(z_{i})}^{2}}
\end{equation}
where $H_{ob}$ denotes observed values, $H_{th}$ denotes the theoretical values of $H(z)$, and $N$ denotes the number of considered datasets of $H(z)$. The $\sigma_{H(z_{i})}$ depicts the standard deviations in $H(z)$ and $\phi=(H_{0}, \Omega_{m0}, \Omega_{r0})$.\\
\begin{figure}[H]
	\centering
	\includegraphics[width=10cm,height=10cm,angle=0]{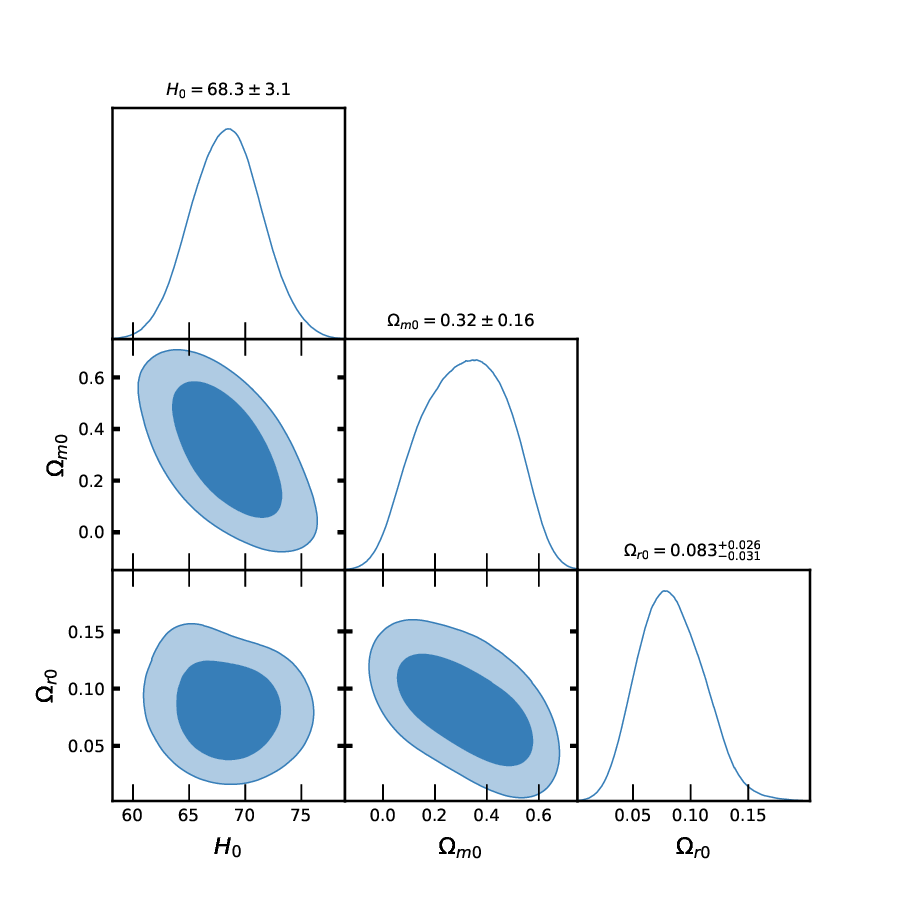}
	\caption{The contour plots of $H_{0}, \Omega_{m0}, \Omega_{r0}$ in MCMC analysis of CC datasets.}
\end{figure}
\begin{table}[H]
	\centering
	\begin{tabular}{|c|c|c|}
		\hline
		
		Parameter      & Prior            & Value                        \\
		\hline
		$H_{0}$        & $(40, 100)$      & $68.3\pm3.1$                 \\
		$\Omega_{m0}$  & $(0, 1)$         & $0.32\pm0.16$                \\
		$\Omega_{r0}$  & $(0, 0.6)$         & $0.083_{-0.031}^{+0.026}$  \\
		$\chi^{2}$     & --               & $14.7359$                    \\
		$\chi^{2}_{red}$& --              & $0.5263$                     \\
		\hline
	\end{tabular}
	\caption{The MCMC Results in $H(z)$ datasets analysis.}\label{T1}
\end{table}
Figure 1 depicts the contour plots of $H_{0}$, $\Omega_{m0}$, and $\Omega_{r0}$ for their best-fit values with CC datasets. The obtained best-fit values of these parameters, $H_{0}$, $\Omega_{m0}$, and $\Omega_{r0}$, are listed in Table 1 with a reduced $\chi^{2}$-value. We found that the present value of the Hubble constant is $H_{0}=68.3\pm3.1$ Km/s/Mpc. The present value of the energy density parameters is $\Omega_{m0}=0.32\pm0.16$ and $\Omega_{r0}=0.083_{-0.031}^{+0.026}$ for the reduced $\chi^{2}_{red}=0.5263$ in the $\sigma-1$ and $\sigma-2$ confidence level.\\

\subsection{Apparent Magnitude $m(z)$}

The correlation between luminosity distance and redshift is a fundamental observational method employed to monitor the progression of the cosmos. When determining the luminosity distance ($D_{L}$) in relation to the cosmic redshift ($z$), the expansion of the universe and the redshift of light from distant bright objects are considered. It is given as
\begin{equation}\label{eq33}
	D_{L}=a_{0} r (1+z),
\end{equation}
where the radial coordinate of the source $r$, is established by
\begin{equation}\label{eq34}
	r  =  \int^r_{0}dr = \int^t_{0}\frac{cdt}{a(t)} = \frac{1}{a_{0}}\int^z_0\frac{cdz'}{H(z')},
\end{equation}
where we have used $ dt=dz/\dot{z}, \dot{z}=-H(1+z)$.\\
Consequently, the following formula determines the luminosity distance:
\begin{equation}\label{eq35}
	D_{L}=c(1+z)\int^z_0\frac{dz'}{H(z')}.
\end{equation}
Supernovae (SNe) are commonly employed by researchers as standard candles to investigate the pace of cosmic expansion using the reported apparent magnitude ($m_{o}$). The surveys on supernovae, which discovered many types of supernovae of varying magnitudes, resulted in the compilation of the Pantheon sample SNe dataset. This dataset has $1048$ data points within the range of $0.01$ to $2.26$ in terms of redshift ($z$). The theoretical apparent magnitude ($m$) of these standard candles is precisely defined as \cite{ref133}.
\begin{equation}\label{eq36}
	m(z)=M+ 5~\log_{10}\left(\frac{D_{L}}{Mpc}\right)+25.
\end{equation}
where $M$ represents the absolute magnitude. The luminosity distance is quantified in units of distance. The Hubble-free luminosity distance ($d_{L}$) can be expressed as $d_{L}\equiv\frac{H_{0}}{c}D_{L}$, where $D_{L}$ is a dimensionless quantity based on $D_{L}$. Therefore, we can express $m(z)$ in a simplified form as shown below
\begin{equation}\label{eq37}
	m(z)=M+5\log_{10}{d_{L}}+5\log_{10}\left(\frac{c/H_{0}}{Mpc}\right)+25.
\end{equation}
The equation above allows for the observation of the degeneracy between $M$ and $H_{0}$, which remains constant in the $\Lambda$CDM background \cite{ref133,ref134}. By redefining, we can combine these deteriorated parameters.
\begin{equation}\label{eq38}
	\mathcal{M}\equiv M+5\log_{10}\left(\frac{c/H_{0}}{Mpc}\right)+25.
\end{equation}
The dimensionless parameter $\mathcal{M}$ is defined as $\mathcal{M}=M-5\log_{10}(h)+42.39$, where $H_{0}$ is the value of $h$ multiplied by $100$ Km/s/Mpc.In the Markov Chain Monte Carlo (MCMC) analysis, we utilize this parameter in conjunction with the appropriate $\chi^{2}$ value for the Pantheon data, as provided in reference \cite{ref134}.
\begin{equation}\nonumber
	\chi^{2}_{P}=V_{P}^{i}C_{ij}^{-1}V_{P}^{j}
\end{equation}
The formula $V_{P}^{i}$ is defined as the subtraction of $m_{o}(z_{i})$ from $m(z)$. The matrix $C_{ij}$ represents the inverse of the covariance matrix, while the value of $m(z)$ is given by Equation \eqref{eq37}.
\begin{table}[H]
	\centering
	\begin{tabular}{|c|c|c|}
		\hline
		Parameter      & Prior      & Value                      \\
		\hline
		$\mathcal{M}$  & $(23, 24)$ & $23.810\pm0.012$           \\
		$\Omega_{m0}$  & $(0, 1)$   & $0.28_{-0.14}^{+0.12}$     \\
		$\Omega_{r0}$  & $(0, 0.6)$ & $0.087_{-0.032}^{+0.043}$  \\
		$\chi^{2}$     & --         & $1026.3478$                \\
		$\chi^{2}_{red}$& --        & $0.9822$                   \\
		\hline
	\end{tabular}
	\caption{The MCMC Results in Pantheon SNe Ia datasets analysis.}\label{T2}
\end{table}
\begin{figure}[H]
	\centering
	\includegraphics[width=10cm,height=10cm,angle=0]{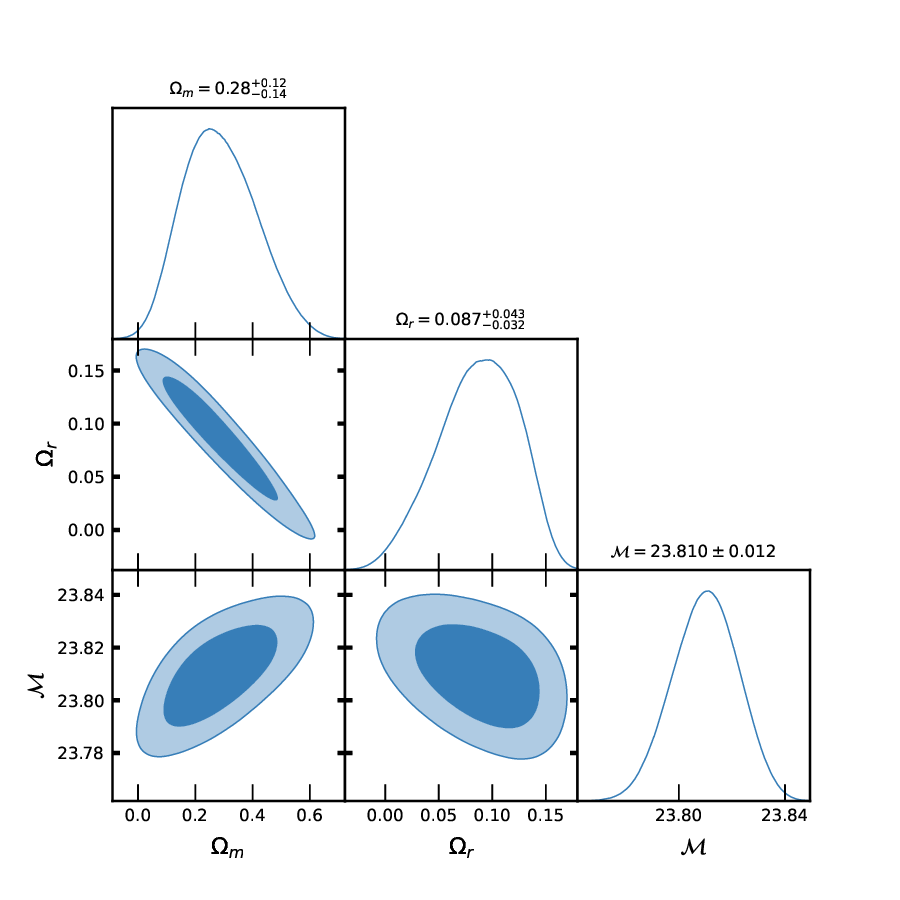}
	\caption{The contour plots of $\Omega_{m0}, \Omega_{r0}, \mathcal{M}$ in MCMC analysis of Pantheon datasets.}
\end{figure}
Figure 2 shows the contour plots of $\Omega_{m0}$, $\Omega_{r0}$, and $\mathcal{M}$ for their best-fit values with Pantheon SNe Ia datasets. Table 2 shows the best-fit values with priors used in MCMC analysis. We found the values of these parameters as $\Omega_{m0}=0.28_{-0.14}^{+0.12}$, $\Omega_{r0}=0.087_{-0.032}^{+0.043}$, and $\mathcal{M}=23.810\pm0.012$ at the reduced $\chi^{2}_{red}=0.9822$ in the $\sigma-1$ and $\sigma-2$ regions of confidence level.

\section{Result analysis and discussions}

This section deals with a detailed result analysis and discussions of recent findings for the derived cosmological models. In section-4, we have estimated the best fit values of the energy density parameters $\Omega_{m0}$, $\Omega_{r0}$, and the Hubble constant value $H_{0}$ along the two observational datasets, the cosmic chronometer (CC) and Pantheon SNe Ia datasets. Now, we can estimate the best fit values of model parameters $\mu$, $\nu$, and $\rho_{m0}$ for a particular choice of arbitrary constant $p_{0}$. In our derived universe model, we take the value of $p_{0}=0$, and hence, we can find the value of $\rho_{m0}=\frac{3H_{0}^{2}\Omega_{m0}}{8\pi}$, $\mu=\frac{64\pi^{2}\Omega_{r0}}{3H_{0}^{2}\Omega_{m0}^{2}}$, and $\nu=6H_{0}^{2}\Omega_{f0}$ in terms of estimated parameters $H_{0}$, $\Omega_{m0}$, and $\Omega_{r0}$. We found the best fit value for the Hubble constant ($H_{0}=68.3\pm3.1, 71.3\pm0.18$ Km/s/Mpc), the curvature energy density parameter ($\Omega_{m0}=0.32\pm0.16, 0.28_{-0.14}^{+0.12}$), and the radiation energy density parameter ($\Omega_{r0}=0.083_{-0.031}^{+0.026}, 0.087_{-0.032}^{+0.043}$) using the CC data and Pantheon datasets. Using these values of $H_{0}$, $\Omega_{m0}$, and $\Omega_{r0}$, we have estimated the present value of energy density $\rho_{m0}=1.8714\times10^{-37}, 1.7845\times 10^{-37} $ gram/cm$^3$, value of model parameters $\mu=3.4829\times10^{37}, 4.3760\times10^{37}$ cm$^3$/gram, and $\nu=1.7549\times10^{-37}, 2.0278\times10^{-37}$ gram/cm$^3$, respectively, along two datasets, CC data and Pantheon datasets. For these best fit values, we have plotted the various cosmological parameters with respect to cosmic redshift $z$ to understand the cosmic history of the expanding universe.\\

We derive the expression for effective EoS parameter $\omega_{eff}$ in Eq.~\eqref{eq28} in terms of energy density parameters and redshift $z$. The variation of $\omega_{eff}$ over $z$ is depicted in figure 3. From Eq.~\eqref{eq28}, one can easily obtain the value of $\omega_{eff}\to\frac{1}{3}$ as $z\to\infty$ (in the early stage) and $\omega_{eff}\to-1$ as $z\to-1$ (in the late stage). This behavior of $\omega_{eff}$ depicts that the early universe was radiation-dominated, and in late-time it became cool down. We found the present value of the effective EoS parameter to be $\omega_{eff}(z=0)=-0.68$ along the CC datasets and $\omega_{eff}(z=0)=-0.70$ for the Pantheon datasets. The latest results from DESI in \cite{ref135} show that $\omega_{0}>-1$ when the equation of state is written as $\omega(a)=\omega_{0}+\omega_{a}(1-a)$. These results show that the current universe model acts as a $\omega_{0}$$\omega_{a}$CDM model. Thus, one can see that our estimated results are compatible with this range, and hence, the derived model acts as a $\omega_{0}$$\omega_{a}$CDM model at present. The present universe model is not the favored $\Lambda$CDM model, which is analyzed in \cite{ref136,ref137}. In general, one can calculate the present value of $\omega_{eff}$ from the Eq.~\eqref{eq28} by putting $z=0$ as $\omega_{eff}=-1+\frac{2}{3}(\Omega_{m0}+2\Omega_{r0})$. Figure 3 shows that the evolution of $\omega_{eff}$ began in a radiation-dominated era with a value of $\frac{1}{3}$. It then went through the structure formation stage with a value of $\omega_{eff}=0$, the transition line with a value of $\omega_{eff}=-\frac{1}{3}$, and the quintessence stage with a value of $-1<\omega_{eff}<-\frac{1}{3}$. Finally, it reached the cosmological constant value of $\omega_{eff}=-1$ in the late-time (at $z=-1$) $\Lambda$CDM universe. These behaviors of $\omega_{eff}$ reveal the consistency of our derived universe model with the observational universe.

\begin{figure}[H]
	\centering
	\includegraphics[width=10cm,height=8cm,angle=0]{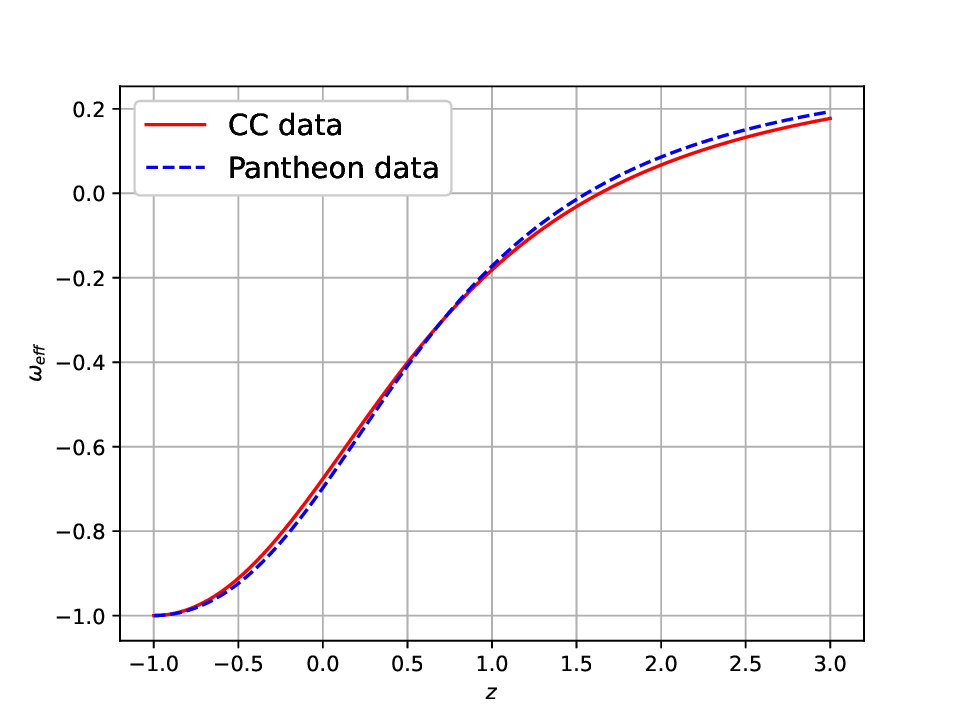}
	\caption{The plot of effective EoS parameter $\omega_{eff}$ versus $z$.}
\end{figure}
We derive the expression for the dimensionless deceleration parameter $q(z)$ as in Eq.~\eqref{eq29}, and its geometrical sketch is given in figure 4. We have estimated the present value of the deceleration parameter $q(z=0)=-0.514$ for the CC datasets and $q(z=0)=-0.546$ for the Pantheon datasets. Recently, using modern cosmography theory \cite{ref138}, the current value of the deceleration parameter was found to be $q_{0}=-0.6633_{-0.6580}^{+0.5753}$ in \cite{ref139}, $q_{0}=-0.528_{-0.088}^{+0.092}$ in \cite{ref140}, $q_{0}=-0.562_{-0.017}^{+0.017}$ in \cite{ref141}, and $q_{0}=-0.59_{-0.03}^{+0.03}$ in \cite{ref142}. As a result, our estimated values of the deceleration parameter $q(z=0)$ are compatible with these recent observed values, revealing the derived model's consistency with the observed universe. Figure 4 shows that $q(z)$ evolves with a positive value and reaches a negative value with decreasing redshift $z$, indicating the transit phase evolution of the expanding universe. At $z=0$, we obtain $q(z=0)=\Omega_{r0}-\Omega_{f0}$, and this implies that $q_{0}>0$ (decelerating phase) for $\Omega_{r0}>\Omega_{f0}$ and $q_{0}<0$ (accelerating phase) for $\Omega_{r0}<\Omega_{f0}$. The transition redshift $z_{t}$ can be obtained from Eq.~\eqref{eq29} for $q(z_{t})=0$ as given below:
\begin{equation}\label{eq39}
	z_{t}=\left(\frac{\Omega_{f0}}{\Omega_{r0}}\right)^{\frac{1}{4}}-1
\end{equation}
The transition redshift $z_{t}$ is estimated as $z_{t}=0.6377$ for CC datasets and $z_{t}=0.6424$ for Pantheon datasets. We observe that $\omega_{eff}(z_{t})=-\frac{1}{3}$, i.e., at $z=z_{t}$ (the transition line), we obtain $q=0$ and $\omega_{eff}=-\frac{1}{3}$. A recent study found that the redshift for the cosmological deceleration to acceleration transition in $f(R)$ gravity is $z_{t}=0.8596_{-0.2722}^{+0.2886}$ for SNIa datasets and $z_{t}=0.6320_{-0.1403}^{+0.1605}$ for Hubble datasets \cite{ref143}. A transition redshift $z_{t}=0.643_{-0.030}^{+0.034}$ in the framework of $f(T)$ gravity is reported in \cite{ref144}. Using a model independent technique, \cite{ref145} measured this transition value as $z_{t}=0.646_{-0.158}^{+0.020}$; \cite{ref146} found it as $z_{t}=0.702_{-0.044}^{+0.094}$; and \cite{ref147,ref148} measured this transition redshift value as $z_{t}=0.684_{-0.092}^{+0.136}$. Thus, one can see that our estimated deceleration-acceleration transition redshift values $z_{t}=0.6377, 0.6424$ are compatible with these recent measurements, which reveal the viability of our derived models.
\begin{figure}[H]
	\centering
	\includegraphics[width=10cm,height=8cm,angle=0]{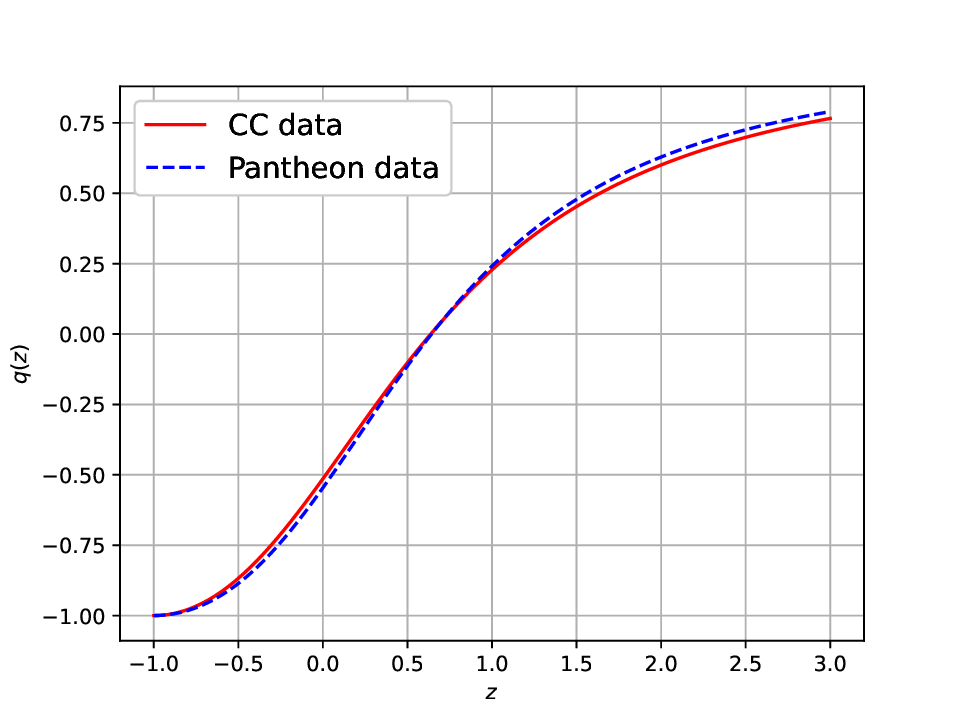}
	\caption{The plot of deceleration parameter $q(z)$ versus $z$.}
\end{figure}
The expression of total energy density parameters for a flat homogeneous and isotropic derived universe is expressed in Eq.~\eqref{eq24} and \eqref{eq30}-\eqref{eq32}, and their geometrical behaviors are represented in figures 5 and 6, respectively, for two observational datasets, CC and Pantheon SNe Ia. We observe that for $z\to\infty$, $(\Omega_{m}, \Omega_{r}, \Omega_{f})\to(0, 1, 0)$ and for $z\to-1$, $(\Omega_{m}, \Omega_{r}, \Omega_{f})\to(0, 0, 1)$ while as $z\to0$, $(\Omega_{m}, \Omega_{r}, \Omega_{f})\to(0.32, 0.083, 0.597)$ along CC datasets and $(\Omega_{m}, \Omega_{r}, \Omega_{f})\to(0.28, 0.087, 0.633)$ along Pantheon datasets.
\begin{figure}[H]
	\centering
	\includegraphics[width=10cm,height=8cm,angle=0]{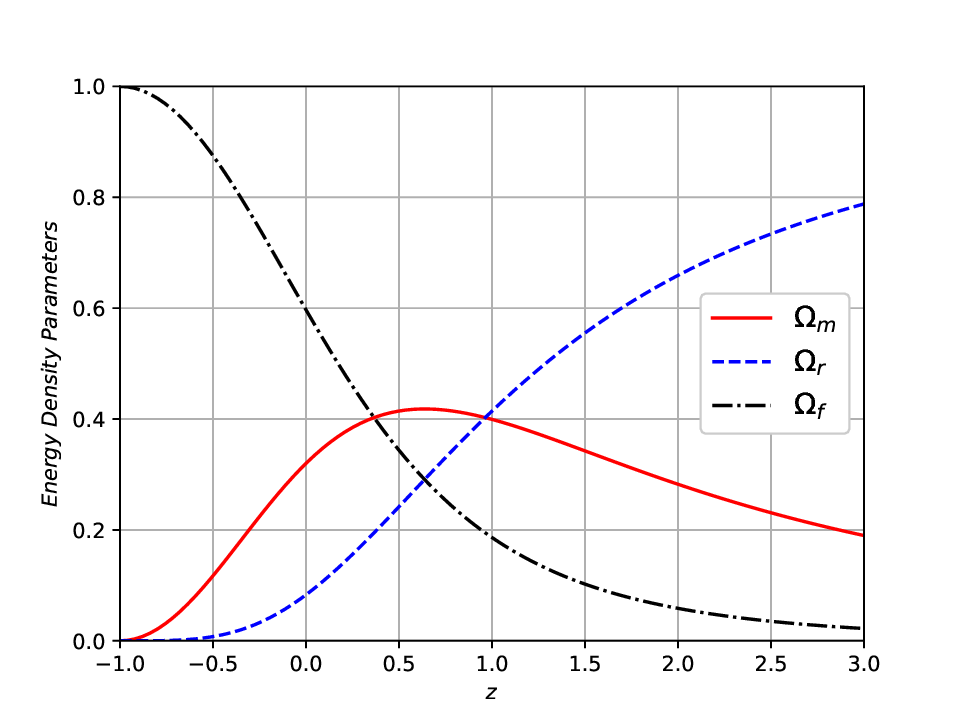}
	\caption{The plot of energy density parameters $\Omega_{m}$, $\Omega_{r}$ and $\Omega_{f}$ versus $z$, respectively for the best fit values of model parameters with CC datasets.}
\end{figure}

\begin{figure}[H]
	\centering
	\includegraphics[width=10cm,height=8cm,angle=0]{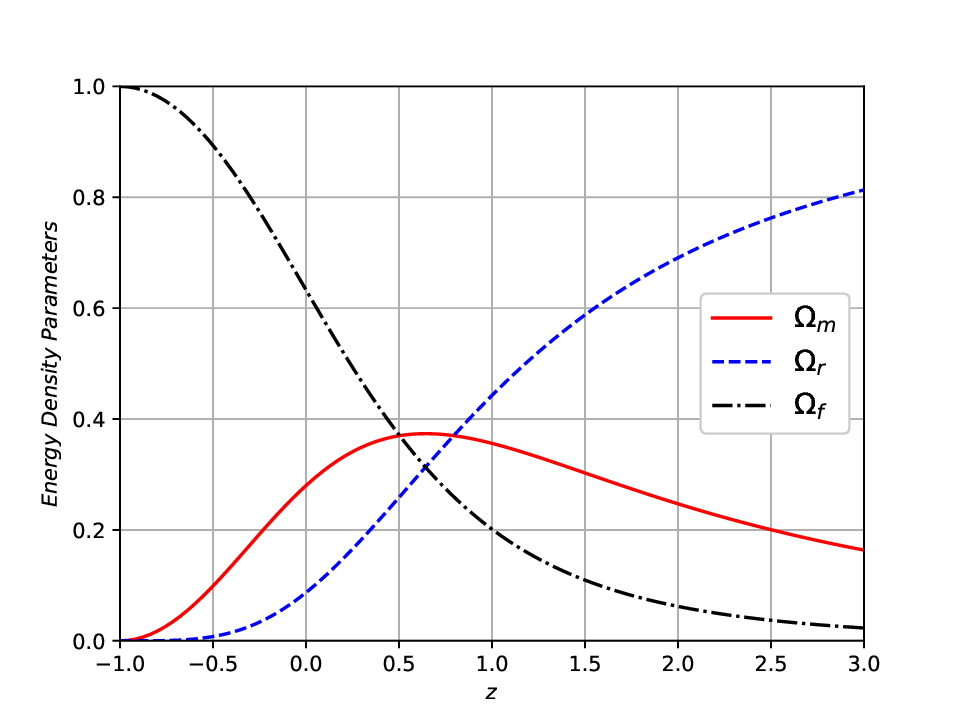}
	\caption{The plot of energy density parameters $\Omega_{m}$, $\Omega_{r}$ and $\Omega_{f}$ versus $z$, respectively for the best fit values of model parameters with Pantheon datasets.}
\end{figure}
Recently, several cosmologists have constrained the value of the Hubble constant $H_{0}$ and energy density parameters from the observational datasets. Cao and Ratra \cite{ref149} estimated the value of matter energy density parameter $\Omega_{m0}=0.288\pm0.017$ with Hubble constant $H_{0}=69.8\pm1.3~Km s^{-1} Mpc^{-1}$. In \cite{ref150}, these quantities are found as $H_{0}=69.7\pm1.2~Km s^{-1} Mpc^{-1}$ and $\Omega_{m0}=0.295\pm0.017$. In \cite{ref151}, these parameters are measured as $H_{0}=66.6\pm1.6~Km s^{-1} Mpc^{-1}$, $\Omega_{m0}=0.29\pm0.02$, while \cite{ref152,ref153} measured as $H_{0}=65.8\pm3.4~Km s^{-1} Mpc^{-1}$, $\Omega_{m0}=0.309\pm0.032$. Recently, the value of Hubble constant was measured as $H_{0}=69.6\pm0.8~Km s^{-1} Mpc^{-1}$ by \cite{ref154} and $H_{0}=67.4_{-3.2}^{+4.1}~Km s^{-1} Mpc^{-1}$ in \cite{ref155}, $H_{0}=69_{-2.8}^{+2.9}~Km s^{-1} Mpc^{-1}$ in \cite{ref156}. Most recently, researchers in \cite{ref157} measured $H_{0}=69.8\pm0.6~Km s^{-1} Mpc^{-1}$, and in \cite{ref158}, it was found as $H_{0}=68.81_{-4.33}^{+4.99}~Km s^{-1} Mpc^{-1}$. According to the Planck collaboration 2018 \cite{ref159}, it is suggested as $H_{0}=67.4\pm0.5$ km/s/Mpc, whereas in 2021 Riess et al. \cite{ref160} found it as $H_{0}=73.2\pm1.3$ km/s/Mpc. Thus, our estimated values of Hubble constant $H_{0}$ and matter energy density parameter $\Omega_{m0}$ is compatible with the value estimated from observational datasets.

\section{Conclusions}

    This study investigated restricted transit cosmological models within the latest suggested modified gravity theory, $f(R,L_{m},T)$-gravity. The modified field equations for a flat, homogeneous, and isotropic Friedmann-Lema\^{\i}tre-Robertson-Walker (FLRW) spacetime metric are derived. The equation of continuity is restricted by enforcing the equation of state for the perfect fluid source, where $p=-\frac{1}{3}\rho+p_{0}$. This allows us to obtain the expression $\dot{\rho}+3H(\rho+p)=0$. The violation of energy conservation is a common occurrence in $f(R,L_{m},T)$-gravity. By applying this constraint, we obtain a connection between the energy density parameters $\Omega_{m0}$, $\Omega_{r0}$, and $\Omega_{f0}$ and the Hubble function. Subsequently, we imposed observational constraints on $H(z)$ in order to determine the most accurate current values of $\Omega_{m0}$, $\Omega_{r0}$, and $H_{0}$. Next, we investigate the cosmological parameters, such as the effective equation of state $\omega_{eff}$, the deceleration parameter, and the energy density parameters $\Omega_{m}$, $\Omega_{r}$, and $\Omega_{f}$. By using these optimal values of energy parameters, we explore the characteristics of various phenomena in the expanding universe.\\
    
    The derived model evolves through a radiation-dominated phase, transitions into a matter-dominated era, and eventually reaches a dark energy-dominated stage (refer to figures 5 and 6). This is one of the model's excellent features. The matter energy density parameter was found to be $\Omega_{m0}=0.32\pm0.16, 0.28_{-0.14}^{+0.12}$, and the Hubble constant was found to be $H_{0}=68.3\pm3.1, 68.3\pm3.1$ Km/s/Mpc. These numbers were found using the CC Hubble data and the Pantheon SNe Ia datasets. These estimated values are compatible with recent observed data in \cite{ref149,ref150,ref151,ref152,ref153,ref154,ref155,ref156,ref157,ref158,ref159,ref160}. we have estimated the present value of energy density $\rho_{m0}=1.8714\times10^{-37}, 1.7845\times 10^{-37} $ gram/cm$^3$, value of model parameters $\mu=3.4829\times10^{37}, 4.3760\times10^{37}$ cm$^3$/gram, and $\nu=1.7549\times10^{-37}, 2.0278\times10^{-37}$ gram/cm$^3$, respectively, along two datasets, CC data and Pantheon datasets.\\    
    
    We made a graph of the state parameter $\omega_{eff}$, which shows the phase of the matter source in the universe, and the dimensionless parameter $q(z)$, which shows the phase of the expansion of the universe, based on the estimated present values of these cosmological parameters. We found the present value of the effective EoS parameter $\omega_{eff}=-0.68, -0.70$ along two datasets, and its range over $-1\le z \le \infty$ is $-1\le\omega_{eff}\le\frac{1}{3}$ (see figure 3), which covers the various matter stages and tends to $\Lambda$CDM model at a late time. Also, an interesting feature of the model is the compatibility of the characteristics of $\omega_{eff}$ with the deceleration parameter $q(z)$. We found the current value of the deceleration parameter as $q_{0}=-0.514, -0.546$ along two observational datasets, respectively, which depicts the current accelerating phase of the universe's expansion. The whole evolution of the $q(z)$ is obtained in the range $(-1, 1)$ over the redshift range $(-1, \infty)$ that depicts the early decelerating to late-time accelerating scenarios of the expanding universe. For $q=0$, the transit line $z=z_{t}$ yields $z_{t}=0.6377$ for CC data and $z_{t}=0.6424$ for Pantheon data. Thus, the derived model represents a transit phase accelerating dark energy model, excluding the concept of a cosmological constant. Therefore, the theory of $f(R, L_{m}, T)$-gravity needs more investigation to explore the observable properties of the universe.

\section*{Acknowledgments}
 
    The author is thankful to the renowned reviewer and editors for their valuable suggestions to improve this manuscript. The author is thankful to IUCAA Center for Astronomy Research and Development (ICARD), CCASS, GLA University, Mathura, India for providing facilities and support where part of this work is carried out.
   
\section{Declarations}
\subsection*{Funding and/or Conflicts of interests/Competing interests}
    The author of this article has no conflict of interests. The author have no competing interests to declare that are relevant to the content of this article. The author did not receive support from any organization for the submitted work.



\begin{thebibliography}{}
	\bibitem {ref1}
	A. G. Riess \textit{et al.}, Observational evidence from supernovae for an accelerating universe and a cosmological constant, \textit{Astron. J.} \textbf{116}, 1009 (1998).
	\bibitem {ref2}
	S. Perlmutter \textit{et al.}, Measurements of Omega and Lambda from 42 High-Redshift Supernovae, \textit{Astrophys. J.} \textbf{517}, 565 (1999).
	\bibitem {ref3}
	R. A. Knop \textit{et al.}, New Constraints on $\Omega_{M}$, $\Omega_{\Lambda}$, and $\omega$ from an Independent Set of $11$ High-Redshift Supernovae Observed with the Hubble Space Telescope, \textit{Astrophys. J.} \textbf{598}, 102 (2003).
	\bibitem {ref4}
	R. Amanullah \textit{et al.}, Spectra and Hubble Space Telescope light curves of six type Ia supernovae at $0.511< z< 1.12$ and the Union2 compilation, \textit{Astrophys. J.} \textbf{716}, 712 (2010).
	\bibitem {ref5}
	D. H. Weinberg \textit{et al.}, Observational probes of cosmic acceleration, \textit{Phys. Rep.} \textbf{530}, 87 (2013).
    \bibitem {ref6}
    A. Einstein, The Principle of General Relativity, \textit{Annalen Physik Leipzig} \textbf{55} (1918) 241-245.
	\bibitem {ref7}
	P. Salucci, N. Turini, and C. Di Paolo, Paradigms and scenarios for the dark matter phenomenon, \textit{Universe} \textbf{6}, 118 (2020).
	\bibitem {ref8}
	S. Alam \textit{et al.} (BOSS Collaboration), The clustering of galaxies in the completed SDSS-III Baryon Oscillation Spectroscopic Survey: cosmological analysis of the DR12 galaxy sample, \textit{Mon. Not. R. Astron. Soc.} \textbf{470}, 2617 (2017). [arXiv:1607.03155].
	\bibitem {ref9}
	T. M. C. Abbott \textit{et al.} (DES Collaboration), Dark Energy Survey year $1$ results: Cosmological constraints from galaxy clustering and weak lensing, \textit{Phys. Rev. D} \textbf{98}, 043526 (2018).
	\bibitem {ref10}
	M. Tanabashi \textit{et al.} (Particle Data Group), Review of Particle Physics: particle data groups, \textit{Phys. Rev. D} \textbf{98}, 030001 (2018).
	\bibitem {ref11}
	N. Aghanim \textit{et al.} (Planck Collaboration), Planck 2018 results. VI. Cosmological parameters, \textit{Astron. Astrophys.} \textbf{641},	A6 (2020).
	\bibitem{ref12}
	E. J. Copeland, M. Sami, S. Tsujikawa, Dynamics of dark energy, \textit{Int. J. Mod. Phys. D} \textbf{15} (2006) 1753-1936. [arXiv:hep-th/0603057v3].
	\bibitem{ref13}
	S. Capozziello, R. D'Agostino, O. Luongo, Thermodynamic parametrization of dark energy, \textit{Phys. Dark Univ.} \textbf{36}, 101045 (2022). [arXiv:2202.03300v2 [astro-ph.CO]].
	\bibitem{ref14}
	P. K. S. Dunsby, O. Luongo, L. Reverberi, Dark Energy and Dark Matter from an additional adiabatic fluid, \textit{Phys. Rev. D} \textbf{94}, 083525 (2016). [arXiv:1604.06908v1 [gr-qc]].
	\bibitem{ref15}
	S. Capozziello, R. D'Agostino, O. Luongo, Extended Gravity Cosmography, \textit{Int. J. Mod. Phys. D} \textbf{28} (2019) 1930016. [arXiv:1904.01427v1 [gr-qc]].
	\bibitem{ref16}
	O. Luongo, H. Quevedo, Characterizing repulsive gravity with curvature eigenvalues, \textit{Phys. Rev. D} \textbf{90} (2014) 084032. [arXiv:1407.1530v2 [gr-qc]].
	\bibitem{ref17}
	O. Luongo, H. Quevedo, Self-accelerated universe induced by repulsive effects as an alternative to dark energy and modified gravities, (2015) [arXiv:1507.06446v1 [gr-qc]].
	\bibitem{ref18}
	A. Aviles, L. Bonanno, O. Luongo, H. Quevedo, Holographic dark matter and dark energy with second order invariants, \textit{Phys. Rev. D} \textbf{84}, 103520, (2011). [arXiv:1109.3177v2 [gr-qc]].
	\bibitem{ref19}
	O. Luongo, M. Muccino, Speeding up the universe using dust with pressure, \textit{Phys. Rev. D} \textbf{98}, 103520 (2018). [arXiv:1807.00180v2 [gr-qc]].
	\bibitem{ref20}
	O. Luongo, M. Muccino, Intermediate redshift calibration of Gamma-ray Bursts and cosmic constraints in non-flat cosmology, \textit{MNRAS} \textbf{518} (2023) 2247-2255. [arXiv:2207.00440v1 [astro-ph.CO]].
	\bibitem {ref21}
	H. A. Buchdahl, Non-linear Lagrangians and cosmological theory, \textit{Mon. Not. R. Astron. Soc.} \textbf{150}, 1 (1970).
	\bibitem {ref22}
	R. Kerner, Cosmology without singularity and nonlinear gravitational Lagrangians, \textit{Gen. Relativ. Gravit.} \textbf{14}, 453 (1982).
	\bibitem {ref23}
	J. P. Duruisseau, R. Kerner, P. Eysseric, Non-Einsteinian gravitational Lagrangians assuring cosmological solutions without collapse, \textit{Gen. Relativ. Gravit.} \textbf{15},	797 (1983).
	\bibitem {ref24}
	J. D. Barrow, A.C. Ottewill, The stability of general relativistic cosmological theory, \textit{J. Phys. A Math. Gen.} \textbf{16}, 2757 (1983).
	\bibitem {ref25}
	H. Kleinert, H.-J. Schmidt, Cosmology with Curvature-Saturated Gravitational Lagrangian R, \textit{Gen. Relativ. Gravit.} \textbf{34}, 1295 (2002).
	\bibitem {ref26}
	S. M. Carroll, V. Duvvuri, M. Trodden, M. S. Turner, Is cosmic speed-up due to new gravitational physics?, \textit{Phys. Rev. D}	\textbf{70}, 043528 (2004).
	\bibitem {ref27}
	W. Hu, I. Sawicki, Models of $f(R)$ cosmic acceleration that evade solar system tests, \textit{Phys. Rev. D} \textbf{76}, 064004 (2007).
	\bibitem {ref28}
	S. A. Appleby, R.A. Battye, Do consistent $F(R)$ models mimic general relativity plus $\Lambda$?, \textit{Phys. Lett. B} \textbf{654}, 7 (2007).
	\bibitem {ref29}
	A. A. Starobinsky, Disappearing cosmological constant in $f(R)$ gravity, \textit{JETP Lett.} \textbf{86}, 157 (2007).
	\bibitem {ref30}
	V. Faraoni, de Sitter space and the equivalence between $f(R)$ and scalar-tensor gravity, \textit{Phys. Rev. D} \textbf{75}, 067302 (2007).
	\bibitem {ref31}
	C. G. B\"{o}hmer, L. Hollenstein, F. S. N. Lobo, Stability of the Einstein static universe in $f(R)$ gravity, \textit{Phys. Rev. D} \textbf{76},	084005 (2007).
	\bibitem {ref32}
	C. G. B\"{o}hmer, T. Harko, F. S. N. Lobo, Dark matter as a geometric effect in $f(R)$ gravity, \textit{Astropart. Phys.} \textbf{29}, 386 (2008).
	\bibitem {ref33}
	C. S. J. Pun, Z. Kovacs, T. Harko, Thin accretion disks in $f(R)$ modified gravity models, \textit{Phys. Rev. D} \textbf{78}, 024043 (2008).
	\bibitem {ref34}
	C. G. B\"{o}hmer, T. Harko, F. S. N. Lobo, The generalized virial theorem in $f(R)$ gravity, \textit{JCAP} \textbf{03}, 024 (2008).
	\bibitem {ref35}
	S. A. Appleby, R. A. Battye, A. A. Starobinsky, Curing singularities in cosmological evolution of $F(R)$ gravity, \textit{JCAP} \textbf{1006}, 005 (2010).
	\bibitem {ref36}
	V. K. Oikonomou, F. P. Fronimos, NTh Chatzarakis, $f(R)$ gravity phase space in the presence of thermal effects, \textit{Phys. Dark Universe} \textbf{30}, 100726 (2020).
	\bibitem {ref37}
	A. V. Astashenok \textit{et al.}, Causal limit of neutron star maximum mass in $f(R)$ gravity in view of GW190814, \textit{Phys. Lett. B} \textbf{816}, 136222 (2021).
	\bibitem {ref38}	
	S. Chakraborty, K. MacDevette, P. Dunsby, Model independent approach to the study of $f(R)$ cosmologies with expansion histories close to $\Lambda$CDM, \textit{Phys. Rev. D} \textbf{103}, 124040 (2021).
	\bibitem {ref39}
	V. K. Oikonomou, Unifying inflation with early and late dark energy epochs in axion $F(R)$ gravity, \textit{Phys. Rev. D} \textbf{103}, 044036 (2021).
	\bibitem {ref40}
	M. A. Mitchell, C. Arnold, B. Li, A general framework to test gravity using galaxy clusters III: observable-mass scaling relations in $f(R)$ gravity, \textit{MNRAS} \textbf{502}, 6101 (2021).
	\bibitem {ref41}
	T. Harko \textit{et al.}, Metric-Palatini gravity unifying local constraints and late-time cosmic acceleration, \textit{Phys. Rev. D} \textbf{85}, 084016 (2012).
	\bibitem {ref42}
	S. Capozziello \textit{et al.}, Hybrid modified gravity unifying local tests, galactic dynamics and late-time cosmic acceleration, \textit{Int. J. Mod. Phys. D} \textbf{22}, 1342006 (2013).
	\bibitem {ref43}
	S. Capozziello \textit{et al.}, Cosmology of hybrid metric-Palatini $f(X)$-gravity, \textit{JCAP} \textbf{04}, 011 (2013).
	\bibitem {ref44}
	S. Capozziello \textit{et al.}, Hybrid metric-Palatini gravity, \textit{Universe} \textbf{1}, 199 (2015).
	\bibitem {ref45}
	Z. Haghani \textit{et al.}, Weyl-Cartan-Weitzenb\"{o}ck gravity as a generalization of teleparallel gravity, \textit{JCAP} \textbf{10}, 061 (2012).
	\bibitem {ref46}
	Z. Haghani \textit{et al.}, Weyl-Cartan-Weitzenb\"{o}ck gravity through Lagrange multiplier, \textit{Phys. Rev. D} \textbf{88}, 044024 (2013).
	\bibitem {ref47}
	J. M. Nester, H.-J. Yo, Symmetric teleparallel general relativity, \textit{Chin. J. Phys.} \textbf{37}, 113 (1999).
	\bibitem {ref48}
	J. Beltran Jimenez, L. Heisenberg, T. Koivisto, Coincident general relativity, \textit{Phys. Rev. D} \textbf{98}, 044048 (2018).
	\bibitem {ref49}
	R. D'Agostino, O. Luongo, Growth of matter perturbations in nonminimal teleparallel dark energy, \textit{Phys. Rev. D} \textbf{98}, 124013 (2018).
	\bibitem {ref50}
	M. Fontanini, E. Huguet, M. Le Delliou, Teleparallel gravity equivalent of general relativity as a gauge theory: Translation or Cartan connection?, \textit{Phys. Rev. D} \textbf{99}, 064006 (2019).
	\bibitem {ref51}
	T. Koivisto, G. Tsimperis, The spectrum of teleparallel gravity, \textit{Universe} \textbf{5}, 80 (2019).
	\bibitem {ref52}
	J. G. Pereira, Y.N. Obukhov, Gauge structure of teleparallel gravity, \textit{Universe} \textbf{5}, 139 (2019).
	\bibitem {ref53}
	D. Blixt, M. Hohmann, C. Pfeifer, On the gauge fixing in the Hamiltonian analysis of general teleparallel theories, \textit{Universe} \textbf{5}, 143 (2019).
	\bibitem {ref54}
	A. A. Coley, R. J. van den Hoogen, D. D. McNutt, Symmetry and equivalence in teleparallel gravity, \textit{J. Math. Phys.} \textbf{61}, 072503 (2020).
	\bibitem {ref55}
	Z. Haghani, N. Khosravi, S. Shahidi, The weyl–cartan gauss–bonnet gravity, \textit{Class. Quantum Gravity} \textbf{32}, 215016 (2015).
	\bibitem {ref56}
	T. P. Sotiriou, V. Faraoni, $f(R)$ theories of gravity, \textit{Rev. Mod. Phys.} \textbf{82}, 451 (2010).
	\bibitem {ref57}
	A. De Felice, S. Tsujikawa, $f(R)$ theories, \textit{Living Rev. Relativ.} \textbf{13}, 3 (2010).
	\bibitem {ref58}
	Y. F. Cai, S. Capozziello, M. De Laurentis, E. N. Saridakis, $f(T)$ teleparallel gravity and cosmology, \textit{Rep. Prog. Phys.} \textbf{79}, 106901 (2016).
	\bibitem {ref59}
	S. Nojiri, S. D. Odintsov, V. K. Oikonomou, Modified gravity theories on a nutshell: Inflation, bounce and late-time evolution, \textit{Phys. Rep.} \textbf{692}, 1 (2017).
	\bibitem {ref60}
	D. Langlois, Dark energy and modified gravity in degenerate higher-order scalar–tensor (DHOST) theories: A review, \textit{Int. J. Mod. Phys. D} \textbf{28}, 1942006 (2019).
	\bibitem {ref61}
	A. Dixit, D. C. Maurya, A. Pradhan, Transit cosmological models coupled with zero-mass scalar field with high redshift in higher derivative theory, \textit{New Astronomy}, \textbf{87}, 101587 (2021).
	\bibitem {ref62}
	A. Pradhan, De Avik, Tee How Loo, D. C. Maurya, A flat FLRW model with dynamical $\Lambda$ as function of matter and geometry, \textit{New Astronomy}, \textbf{89}, 101637 (2021).
	\bibitem {ref63}
	D. C. Maurya, R. Zia, A. Pradhan, Anisotropic string cosmological model in Brans–Dicke theory of gravitation with time-dependent deceleration parameter, \textit{J. Exp. Theor. Phys.}, \textbf{123}, 617-622 (2016).
	\bibitem {ref64}
	D. C. Maurya, R. Zia, A. Pradhan, Dark energy models in LRS Bianchi type-II space-time in the new perspective of time-dependent deceleration parameter, \textit{Int. J. Geom. Meth. Mod. Phys.}, \textbf{14}(05) 1750077 (2017).
	\bibitem {ref65}
	D. C. Maurya, Anisotropic dark energy transit cosmological models with time-dependent $\omega(t)$ and redshift-dependent $\omega(z)$ EoS parameter, \textit{Int. J. Geom. Meth. Mod. Phys.}, \textbf{15}(02) 1850019 (2018).
	\bibitem {ref66}
	R. Zia, D. C. Maurya, Brans–Dicke scalar field cosmological model in Lyra’s geometry with time-dependent deceleration parameter, \textit{Int. J. Geom. Meth. Mod. Phys.}, \textbf{15}(11) 1850186 (2018).
	\bibitem {ref67}
	D. C. Maurya, R. Zia, Brans-Dicke scalar field cosmological model in Lyra’s geometry, \textit{Phys. Rev. D}, \textbf{100}(2) 023503 (2019).
	\bibitem {ref68}
	R. Zia, U.K. Sharma, D.C. Maurya, Transit two-fluid models in anisotropic Bianchi type-III space-time, \textit{New Astronomy}, \textbf{72}, 83-91 (2019).
	\bibitem {ref69}
	D. C. Maurya, R. Zia, Reply to ``Comment on `Brans-Dicke scalar field cosmological model in Lyra's geometry'", \textit{Phys. Rev. D}, \textbf{102}(10) 108302 (2020).	
	\bibitem {ref70}
	O. Bertolami, C. G. Boehmer, T. Harko, F. S. N. Lobo, Extra force in $f(R)$ modified theories of gravity, \textit{Phys. Rev. D} \textbf{75}, 104016 (2007).
	\bibitem {ref71}
	T. Harko, Modified gravity with arbitrary coupling between matter and geometry, \textit{Phys. Lett. B} \textbf{669}, 376 (2008).
	\bibitem {ref72}
	T. Harko, F. S. N. Lobo, $f(R,L_{m})$ gravity, \textit{Eur. Phys. J. C} \textbf{70}, 373 (2010).
	\bibitem {ref73}
	T. Harko, The matter Lagrangian and the energy-momentum tensor in modified gravity with nonminimal coupling between matter and geometry, \textit{Phys. Rev. D} \textbf{81}, 044021 (2010).
	\bibitem {ref74}
	T. Harko, F. S. N. Lobo, Geodesic deviation, Raychaudhuri equation, and tidal forces in modified gravity with an arbitrary curvature-matter coupling, \textit{Phys. Rev. D} \textbf{86}, 124034 (2012).
	\bibitem {ref75}
	J. Wang, K. Liao, Energy conditions in $f(R, L_{m})$ gravity, \textit{Class. Quantum Gravity} \textbf{29}, 215016 (2012).
	\bibitem {ref76}
	O. Minazzoli, T. Harko, New derivation of the Lagrangian of a perfect fluid with a barotropic equation of state, \textit{Phys. Rev. D} \textbf{86}, 087502 (2012).
	\bibitem {ref77}
	T. Harko, F. S. N. Lobo, O. Minazzoli, Extended $f(R,L_{m})$ gravity with generalized scalar field and kinetic term dependences, \textit{Phys. Rev. D} \textbf{87}, 047501 (2013).
	\bibitem {ref78}
	D. W. Tian, I. Booth, Lessons from $(R,R_{c}^{2},R_{m}^{2},L_{m})$ gravity: Smooth Gauss-Bonnet limit, energy-momentum conservation, and nonminimal coupling, \textit{Phys. Rev. D} \textbf{90}, 024059 (2014).
	\bibitem {ref79}
	T. Harko, F. S. N. Lobo, J. P. Mimoso, D. Pav\'{o}n, Gravitational induced particle production through a nonminimal curvature–matter coupling, \textit{Eur. Phys. J. C} \textbf{75}, 386 (2015).
	\bibitem {ref80}
	R. P. L. Azevedo, P. P. Avelino, Big-bang nucleosynthesis and cosmic microwave background constraints on nonminimally coupled theories of gravity, \textit{Phys. Rev. D} \textbf{98}, 064045 (2018).
	\bibitem {ref81}
	S. Bahamonde, Generalised nonminimally gravity-matter coupled theory, \textit{Eur. Phys. J. C} \textbf{78}, 326 (2018).
	\bibitem {ref82}
	R. March, O. Bertolami, M. Muccino, R. Baptista, S.	Dell'Agnello, Constraining a nonminimally coupled curvature-matter gravity model with ocean experiments, \textit{Phys. Rev. D} \textbf{100}, 042002 (2019).
	\bibitem {ref83}
	O. Bertolami, C. Gomes, Nonminimally coupled Boltzmann equation: foundations, \textit{Phys. Rev. D} \textbf{102}, 084051 (2020).
	\bibitem {ref84}
	A. Pradhan, D. C. Maurya, G. K. Goswami, A. Beesham, Modeling Transit Dark Energy in $F(R, L_{m})$-gravity, \textit{Int. J. Geom. Meth. Mod. Phys.}, \textbf{20}(06) 2350105 (2023).
	\bibitem {ref85}
	D. C. Maurya, Accelerating scenarios of massive universe in $f(R, L_{m})$-gravity, \textit{New Astronomy}, \textbf{100}, 101974 (2023).
	\bibitem {ref86}
	D. C. Maurya, Exact Cosmological Models in Modified $F(R,L_{m})$-Gravity with Observational Constraints, \textit{Grav. \& Cosmo.}, \textbf{29}(3) 315-325 (2023).
	\bibitem {ref87}
	D. C. Maurya, Bianchi-I dark energy cosmological model in $f(R,L_{m})$-Gravity, \textit{Int. J. Geom. Meth. Mod. Phys.}, \textbf{21}(04) 2450072-194 (2024).
	\bibitem {ref88}
	D. C. Maurya, Constrained $\Lambda$CDM Dark Energy Models in Higher Derivative $F(R,L_{m})$-Gravity Theory, \textit{Phys. Dark Universe}, \textbf{42} (2023) 101373.
	\bibitem {ref89}
	T. Harko, F. S. N. Lobo, S. Nojiri, S. D. Odintsov, $f(R,T)$ gravity, \textit{Phys. Rev. D} \textbf{84}, 024020 (2011).
	\bibitem {ref90}
	F. G. Alvarenga, A. de la Cruz-Dombriz, M. J. S. Houndjo, M. E.	Rodrigues, D. Saez-Gomez, Dynamics of scalar perturbations in $f(R, T)$ gravity, \textit{Phys. Rev. D} \textbf{87}, 103526 (2013).
	\bibitem {ref91}
	T. Harko, Thermodynamic interpretation of the generalized gravity models with geometry-matter coupling, \textit{Phys. Rev. D} \textbf{90}, 044067 (2014).
	\bibitem {ref92}
	E. H. Baffou, M. J. S. Houndjo, M. E. Rodrigues, A. V. Kpadonou, J. Tossa, Cosmological evolution in $f(R,T)$ theory with collisional matter, \textit{Phys. Rev. D} \textbf{92}, 084043 (2015).
	\bibitem {ref93}
	P. H. R. S. Moraes, J. D. V. Arbanil, M. Malheiro, Stellar equilibrium configurations of compact stars in $f(R, T)$ theory of gravity, \textit{JCAP} \textbf{06}, 005	(2016).
	\bibitem {ref94}
	M. E. S. Alves, P. H. R. S. Moraes, J. C. N. de Araujo, M. Malheiro, Gravitational waves in $f(R,T)$ and $f(R,T^{\phi})$ theories of gravity, \textit{Phys. Rev. D} \textbf{94}, 024032 (2016).
	\bibitem {ref95}
	M. Zubair, S. Waheed, Y. Ahmad, Static spherically symmetric wormholes in $f(R, T)$ gravity, \textit{Eur. Phys. J. C} \textbf{76}, 444 (2016).
	\bibitem {ref96}	
	M.-X. Xu, T. Harko, S.-D. Liang, Quantum Cosmology of $f(R, T)$ gravity, \textit{Eur. Phys. J. C} \textbf{76}, 449 (2016).
	\bibitem {ref97}
	H. Shabani, A.H. Ziaie, Stability of the Einstein static universe in $f(R, T)$ gravity, \textit{Eur. Phys. J. C} \textbf{77}, 31 (2017).
	\bibitem {ref98}
	P. H. R. S. Moraes, P. K. Sahoo, Modeling wormholes in $f(R,T)$ gravity, \textit{Phys. Rev. D} \textbf{96}, 044038 (2017).
	\bibitem {ref99}
	E. H. Baffou, M. J. S. Houndjo, M. Hamani-Daouda, F. G.	Alvarenga, Late-time cosmological approach in mimetic $f(R, T)$ gravity, \textit{Eur. Phys. J. C} \textbf{77}, 708 (2017).
	\bibitem {ref100}
	F. Rajabi, K. Nozari, Unimodular $f(R,T)$ gravity, \textit{Phys. Rev. D} \textbf{96}, 084061 (2017).
	\bibitem {ref101}
	H. Velten, T. R. P. Caram\^{e}s, Cosmological inviability of $f(R,T)$ gravity, \textit{Phys. Rev. D} \textbf{95}, 123536 (2017).
	\bibitem {ref102}
	P. K. Sahoo, P. H. R. S. Moraes, P. Sahoo, Wormholes in $R^{2}$-gravity within the $f(R, T)$ formalism, \textit{Eur. Phys. J. C} \textbf{78}, 46 (2018).
	\bibitem {ref103}
	J. K. Singh, K. Bamba, R. Nagpal, S. K. J. Pacif, Bouncing cosmology in $f(R,T)$ gravity, \textit{Phys. Rev. D} \textbf{97}, 123536 (2018).
	\bibitem {ref104}
	J. Wu, G. Li, T. Harko, S.-D. Liang, Palatini formulation of $f(R, T)$ gravity theory, and its cosmological implications, \textit{Eur. Phys. J. C} \textbf{78}, 430 (2018).
	\bibitem {ref105}
	R. Nagpal, S. K. J. Pacif, J. K. Singh, K. Bamba, A. Beesham, Analysis with observational constraints in $\Lambda$-cosmology in $f(R, T)$ gravity, \textit{Eur.	Phys. J. C} \textbf{78}, 946 (2018).
	\bibitem {ref106}
	P. H. R. S. Moraes, R. A. C. Correa, G. Ribeiro, Evading the non-continuity equation in the $f(R, T)$ cosmology, \textit{Eur. Phys. J. C} \textbf{78}, 192 (2018).
	\bibitem {ref107}
	D. Deb, S. V. Ketov, M. Khlopov, S. Ray, Study on charged strange stars in $f(R, T)$ gravity, \textit{J. Cosmol. Astropart. Phys.} \textbf{10}, 070 (2019).
	\bibitem {ref108}
	P. H. R. S. Moraes, The trace of the trace of the energy–momentum tensor-dependent Einstein's field equations, \textit{Eur. Phys. J. C} \textbf{79}, 674 (2019).
	\bibitem {ref109}
	S. K. Maurya, A. Errehymy, D. Deb, F. Tello-Ortiz, M. Daoud, Study of anisotropic strange stars in $f(R,T)$ gravity: An embedding approach under the simplest linear functional of the matter-geometry coupling, \textit{Phys. Rev. D} \textbf{100}, 044014 (2019).
	\bibitem {ref110}
	S. K. Maurya \textit{et al.}, Gravitational decoupling minimal geometric deformation model in modified $f(R, T)$ gravity theory, \textit{Phys. Dark Universe} \textbf{30}, 100640 (2020).
	\bibitem {ref111}
	R. Lobato \textit{et al.}, Neutron stars in $f(R, T)$ gravity using realistic equations of state in the light of massive pulsars and GW170817, \textit{J. Cosmol. Astropart. Phys.} \textbf{12}, 039 (2020).
	\bibitem {ref112}
	T. Harko, P. H. R. S. Moraes, Comment on ``Reexamining $f(R,T)$ gravity", \textit{Phys. Rev. D} \textbf{101}, 108501 (2020).
	\bibitem {ref113}
	G. A. Carvalho, F. Rocha, H. O. Oliveira, R. V. Lobato, General approach to the Lagrangian ambiguity in $f(R, T)$ gravity, \textit{Eur. Phys. J. C} \textbf{81}, 134 (2021).
	\bibitem {ref114}
	M. Gamonal, Slow-roll inflation in $f(R, T)$ gravity and a modified Starobinsky-like inflationary model, \textit{Phys. Dark Universe} \textbf{31}, 100768 (2021).
	\bibitem {ref115}
	J. M. Z. Pretel, S. E. Jor\'{a}s, R. R. R. Reis, J. D. V. Arba\~{n}il, Radial oscillations and stability of compact stars in $f(R, T)= R+2\beta T$ gravity, \textit{JCAP} \textbf{04}, 064 (2021).
	\bibitem {ref116}
	R. Zia, D. C. Maurya, A. Pradhan, Transit dark energy string cosmological models with perfect fluid in $F(R,T)$-gravity, \textit{Int. J. Geom. Meth. Mod. Phys.}, \textbf{15}(10) 1850168 (2018).
	\bibitem {ref117}
	D. C. Maurya, Modified $F(R,T)$ cosmology with observational constraints in Lyra’s geometry, \textit{Int. J. Geom. Meth. Mod. Phys.}, \textbf{17}(01) 2050001 (2020).
	\bibitem {ref118}
	D. C. Maurya, A. Pradhan, A. Dixit, Domain walls and quark matter in Bianchi type-V universe with observational constraints in $F(R,T)$ gravity, \textit{Int. J. Geom. Meth. Mod. Phys.}, \textbf{17}(01) 2050014 (2020).
	\bibitem {ref119}
	D. C. Maurya, Transit cosmological model with specific Hubble parameter in $F(R,T)$ gravity, \textit{New Astronomy}, \textbf{77}, 101355 (2020).
	\bibitem {ref120}
	D. C. Maurya, J. singh, L. K. Gaur, Dark Energy Nature in Logarithmic $f(R,T)$ Cosmology, \textit{Int. J. Geom. Meth. Mod. Phys.}, \textbf{20}(11) 2350192 (2023).
	\bibitem {ref121}
	D. C. Maurya, R. Myrzakulov, Transit cosmological models in $F(R,\bar{T})$ gravity theory, \textit{Eur. Phys. J. C}, \textbf{84}(5) (2024) 534.
	\bibitem {ref122}
	D. C. Maurya, R. Myrzakulov, Exact cosmological models in metric-affine $F(R, T)$ gravity, \textit{Eur. Phys. J. C}, \textbf{84} (2024) 625.
	\bibitem {ref123}
	G. P. Singh, N. Hulke, A. Singh, Cosmological study of particle creation in higher derivative theory, \textit{Indian J. Phys.} \textbf{94} 127-141 (2020).
	\bibitem {ref124}
	N. Hulke, G. P. Singh, B. K. Bishi, A. Singh, Variable Chaplygin gas cosmologies in $f(R, T)$ gravity with particle creation, \textit{New Astronomy} \textbf{77} 101357 (2020).
	\bibitem {ref125}
	U. K. Sharma, M. Kumar, G. Varshney, Scalar Field Models of Barrow Holographic Dark Energy in $f(R,T)$ Gravity, \textit{Universe} \textbf{8} 642 (2022).
	\bibitem {ref126}
	U. K. Sharma, A. Pradhan, Cosmology in modified $f(R,T)$-gravity theory in a variant $\Lambda(t)$ scenario-revisited, \textit{Int. J. Geom. Meth. Mod. Phys.}, \textbf{15} 1850014 (2018).	
	\bibitem {ref127}
    Z Haghani1 and T. Harko, Generalizing the coupling between geometry and matter: $f(R, L_{m}, T)$ gravity, \textit{Eur. Phys. J. C} (2021) \textbf{81}:615.
    \bibitem{ref128}
    S. Capozziello, \textit{et al.}, Cosmic acceleration in non-flat $f(T)$ cosmology, \textit{Gen. Relativ. Gravit.} \textbf{50}, 53 (2018).  [arXiv:1804.03649v1 [gr-qc]].  
    \bibitem {ref129}
    L. D. Landau, E. M. Lifshitz, The Classical Theory of Fields (Butterworth-Heinemann, Oxford, 1998).    
    \bibitem {ref130}
    D. Foreman-Mackey, D. W. Hogg, D. Lang, J. Goodman, \textit{Publ. Astron. Soc. Pac.} \textbf{125}, 306 (2013). https://doi.org/10.1086/670067
    \bibitem {ref131}
    J. Simon, L. Verde, R. Jimenez, \textit{Phys. Rev. D} \textbf{71}, 123001 (2005). https://doi.org/10.1103/PhysRevD.71.123001
    \bibitem {ref132}
    G. S. Sharov, V. O. Vasiliev, \textit{Math. Model. Geom.} \textbf{6}, 1-20 (2018). https://doi.org/10.26456/mmg/2018-611.
    \bibitem {ref133}
    G. Ellis, R. Maartens, M. MacCallum, Relativistic Cosmology (Cambridge University Press, Cambridge, 2012). https://doi.org/10.1017/CBO9781139014403.
    \bibitem {ref134}
    K. Asvesta, L. Kazantzidis, L. Perivolaropoulos, C.G. Tsagas, \textit{Mon. Not. R. Astron. Soc.} \textbf{513}, 2394-2406 (2022). https://doi.org/10.1093/mnras/stac922.
    \bibitem{ref135}
    A. G. Adame, \textit{et al.}, [DESI Collaboration], DESI 2024 VI: Cosmological Constraints from the Measurements of Baryon Acoustic Oscillations, (2014) [arXiv:2404.03002v2 [astro-ph.CO]].
    \bibitem{ref136}
    Y. Carloni, O. Luongo, M. Muccino, Does dark energy really revive using DESI 2024 data?, (2024) [arXiv:2404.12068v1 [astro-ph.CO]].
    \bibitem{ref137}
    O. Luongo, M. Muccino, Model independent cosmographic constraints from DESI 2024, (2024) [arXiv:2404.07070v1 [astro-ph.CO]].
    \bibitem{ref138}
    P. K. S. Dunsby, O. Luongo, On the theory and applications of modern cosmography, \textit{Int. J. Geom. Meth. Mod. Phys.} \textbf{13} 1630002 (2016). [arXiv:1511.06532v1 [gr-qc]].
     \bibitem{ref139}
    A. Aviles, C. Gruber, O. Luongo, H. Quevedo, Cosmography and constraints on the equation of state of the Universe in various parametrizations, \textit{Phys. Rev. D} \textbf{86} 123516 (2012). [arXiv:1204.2007v2 [astro-ph.CO]].
    \bibitem{ref140}
    C. Gruber, O. Luongo, Cosmographic analysis of the equation of state of the universe through Pad\'{e} approximations, \textit{Phys. Rev. D} \textbf{89}, 103506 (2014). [arXiv:1309.3215v1 [gr-qc]].
    \bibitem{ref141}
    A. Aviles, J. Klapp, O. Luongo, Toward unbiased estimations of the statefinder parameters, (2017) [arXiv:1606.09195v2 [astro-ph.CO]].
    \bibitem{ref142}
    O. Luongo, M. Muccino, Kinematic constraints beyond $z\backsimeq0$ using calibrated GRB correlations, \textit{A \& A} \textbf{641}, A17, (2020). [arXiv:2010.05218v1 [astro-ph.CO]].
    \bibitem{ref143}
    S. Capozziello, O. Farooq, O. Luongo, B. Ratra, Cosmographic bounds on the cosmological deceleration-acceleration transition redshift in $f(R)$ gravity, \textit{Phys. Rev. D} \textbf{90}, 044016 (2014). [arXiv:1403.1421v1 [gr-qc]].
    \bibitem{ref144}
    S. Capozziello, O. Luongo, E. N. Saridakis, Transition redshift in $f(T)$ cosmology and observational constraints, \textit{Phys. Rev. D} \textbf{91} 124037 (2015). [arXiv:1503.02832v2 [gr-qc]].
    \bibitem{ref145}
    S. Capozziello, P. K. S. Dunsby, O. Luongo, Model independent reconstruction of cosmological accelerated-decelerated phase, \textit{MNRAS} \textbf{509} (2022) 5399-5415. [arXiv:2106.15579v2 [astro-ph.CO]].
    \bibitem{ref146}
    M. Muccino, O. Luongo, D. Jain, Constraints on the transition redshift from the calibrated Gamma-ray Burst Ep-Eiso correlation, \textit{MNRAS} \textbf{523} (2023) 4938-4948. [arXiv:2208.13700v3 [astro-ph.CO]].
    \bibitem{ref147}
    A. C. Alfano, S. Capozziello, O. Luongo, M. Muccino, Cosmological transition epoch from gamma-ray burst correlations, (2024) [arXiv:2402.18967v1 [astro-ph.CO]].
    \bibitem{ref148}
    A. C. Alfano, C. Cafaro, S. Capozziello, O. Luongo, Dark energy-matter equivalence by the evolution of cosmic equation of state, \textit{Phys. Dark Univ.} \textbf{42} 101298 (2023). [arXiv:2306.08396v2 [astro-ph.CO]]. 
    \bibitem {ref149}
    S. Cao and B. Ratra, $H_{0}=69.8\pm1.3~km~s^{-1}~Mpc^{-1}$, $\Omega_{m0}=0.288\pm0.017$, and other constraints from lower-redshift, non-CMB, expansion-rate data, \textit{Phys. Rev. D} \textbf{107}, 103521 (2023). [arXiv:2302.14203 [astro-ph.CO]].
    \bibitem {ref150}
    S. Cao and B. Ratra, Using lower-redshift, non-CMB, data to constrain the Hubble constant and other cosmological parameters, \textit{MNRAS} \textbf{513}, 5686-5700 (2022). [arXiv:2203.10825 [astro-ph.CO]].
    \bibitem {ref151}
    A. Dom\'{\i}nguez \textit{et al.}, A new measurement of the Hubble constant and matter content of the Universe using extra-galactic background light $\gamma$-ray attenuation, (2019) [arXiv:1903.12097v2 [astro-ph.CO]].
    \bibitem {ref152}
    Chan-Gyung Park, Bharat Ratra, Using SPTpol, Planck 2015, and non-CMB data to constrain tilted spatially-flat and untilted non-flat $\Lambda$CDM, XCDM, and $\phi$CDM dark energy inflation cosmologies, \textit{Phys. Rev. D} \textbf{101}, 083508 (2020). [arXiv:1908.08477 [astro-ph.CO]].
    \bibitem {ref153}
    W. Lin and M. Ishak, A Bayesian interpretation of inconsistency measures in cosmology, \textit{JCAP} \textbf{2105}:009, (2021). [arXiv:1909.10991v3 [astro-ph.CO]].
    \bibitem {ref154}
    W. L. Freedman \textit{et al.}, Calibration of the Tip of the Red Giant Branch (TRGB), (2020) [arXiv:2002.01550v1 [astro-ph.GA]].
    \bibitem {ref155}
    S. Birrer \textit{et al.}, TDCOSMO IV: Hierarchical time-delay cosmography -- joint inference of the Hubble constant and galaxy density profiles, \textit{A \& A} \textbf{643}, A165 (2020). [arXiv:2007.02941v3 [astro-ph.CO]].
    \bibitem {ref156}
    Supranta S. Boruah, Michael J. Hudson, Guilhem Lavaux, Peculiar velocities in the local Universe: comparison of different models and the implications for $H_{0}$ and dark matter, (2020) [ arXiv:2010.01119v1 [astro-ph.CO] ].
    \bibitem {ref157}
    Wendy L. Freedman, Measurements of the Hubble Constant: Tensions in Perspective, (2021). arXiv:2106.15656v1 [astro-ph.CO].
    \bibitem {ref158}
    Q. Wu, G. Q. Zhang, F. Y. Wang, An $8\%$ Determination of the Hubble Constant from localized Fast Radio Bursts, (2022) [arXiv:2108.00581v2 [astro-ph.CO] ].
    \bibitem{ref159}
    Planck Collaboration, Aghanim N., Akrami Y. \textit{et al.} (2020) Planck 2018 results. VI. Cosmological
    parameters. A \& A 641:A6. https://doi.org/10.1051/0004-6361/201833910. arXiv:1807.06209 [astroph.CO]
    \bibitem{ref160}
    Riess A. G., Casertano S., Yuan W. \textit{et al.} (2021) Cosmic distances calibrated to $1\%$ precision with Gaia
    EDR3 parallaxes and Hubble Space Telescope photometry of 75 Milky Way Cepheids Confirm Tension with KCDM. ApJ 908(1):L6. https://doi.org/10.3847/20418213/abdbaf. arXiv:2012.08534
    [astro-ph.CO]
\end{thebibliography}
\end{document}